\documentclass[10pt,journal]{IEEEtran}
\ifCLASSINFOpdf
\else
\fi
\usepackage{optidef}
\usepackage{etoolbox}
\usepackage{bm}
\patchcmd{\bibsetup}{\interlinepenalty=5000}{\interlinepenalty=0}{}{}
\allowbreak
\usepackage{titlesec}
\usepackage[skip=0pt]{caption} 
\allowdisplaybreaks
\usepackage{braket}
\usepackage{lettrine}

\usepackage{etoolbox}
\appto{\bibsetup}{\raggedbottom}

\usepackage[labelformat=simple]{subcaption}  

\usepackage{algorithm}
\usepackage{placeins}
\usepackage{comment}
\usepackage{algpseudocode}
\usepackage[english]{babel}

\usepackage{mathtools}
\usepackage{soul}
\usepackage{pdfpages}

\usepackage[skip=3pt]{caption}

\usepackage{etoolbox}
\AfterEndEnvironment{algorithm}{\vspace{-3pt}}

\AtBeginDocument{%
  \setlength\abovedisplayskip{6pt}
  \setlength\belowdisplayskip{6pt}
  \setlength\abovedisplayshortskip{3pt}
  \setlength\belowdisplayshortskip{3pt}
}

\usepackage{titlesec}
\titlespacing*{\section}{0pt}{6pt plus 2pt minus 1pt}{3pt plus 1pt minus 1pt}
\titlespacing*{\subsection}{0pt}{5pt plus 1pt}{2pt plus 1pt}
\titlespacing*{\subsubsection}{0pt}{4pt plus 1pt}{2pt plus 1pt}

\makeatletter
\def\thm@space@setup{%
  \thm@preskip=3pt
  \thm@postskip=3pt
}
\makeatother
\newtheorem{theorem}{Theorem}[section]
\newtheorem{lemma}{Lemma}

\newtheorem{proof2}{Proof}

\newtheorem{example}[theorem]{Example}
\newtheorem{remark}[theorem]{Remark}

\usepackage{tabularx,booktabs}
\usepackage {xcolor}
\usepackage[utf8]{inputenc}
\usepackage{amssymb}
\usepackage{etoolbox}
\AtBeginDocument{%
  \setlength\abovedisplayskip{6pt}
  \setlength\belowdisplayskip{6pt}
  \setlength\abovedisplayshortskip{3pt}
  \setlength\belowdisplayshortskip{3pt}
}
\usepackage{color}
\usepackage{graphicx,amssymb,amsfonts,booktabs,multirow,array,comment}

\usepackage[all,graph]{xy}

\makeatletter
\def\thm@space@setup{%
  \thm@preskip=0pt
  \thm@postskip=0pt
}
\makeatother

\usepackage{mathtools}

\appto{\bibsetup}{\raggedbottom}
\begin{document}
\bstctlcite{IEEEexample:BSTcontrol}

\hyphenation{op-tical net-works semi-conduc-tor}

\title{Path-Based Quantum Meta-Learning for Adaptive Optimization of Reconfigurable Intelligent Surfaces}
\author{Noha~Hassan,~\IEEEmembership{SMIEEE,}
        Xavier~Fernando,~\IEEEmembership{SMIEEE,}
        and~Halim~Yanikomeroglu,~\IEEEmembership{FIEEE}
\thanks{Submitted to IEEE Wireless Communications Letters; access may be restricted upon copyright transfer.}
\thanks{N. Hassan and H. Yanikomeroglu are with the Department of Systems and Computer Engineering,
Carleton University, Ottawa, ON, Canada
(e-mail:  noha.hassan@torontomu.ca; halim@sce.carleton.ca).}
\thanks{X. Fernando is with the Department of Electrical, Computer, and Biomedical Engineering,
Toronto Metropolitan University, Toronto, ON, Canada
(e-mail: fernando@torontomu.ca).}}

\maketitle

\markboth{IEEE Wireless Communications Letters}%
{Hassan \MakeLowercase{\textit{et al.}}: Path-Based Quantum Meta-Learning for RIS}

\pagestyle{headings}

\vspace{-40pt}

\begin{abstract}
Reconfigurable intelligent surfaces (RISs) modify signal reflections to enhance wireless communication capabilities. Classical RIS phase optimization is highly non convex and challenging in dynamic environments due to high interference and user mobility. 

Here we propose a hierarchical multi-objective \textit{quantum} meta-learning algorithm that switches among specific quantum paths based on historical success, energy cost, and current data rate.  Candidate RIS control directions are arranged as switch paths between quantum neural network layers to minimize inference, and a scoring mechanism selects the top performing paths per layer. Instead of merely storing past successful settings of the RIS and picking the closest match when a new problem is encountered, the algorithm learns how to select and recombine the best parts of different solutions to solve new scenarios. 
In our model, high-dimensional RIS scenario features are compressed into a quantum state using the tensor product, then superimposed during quantum path selection, significantly improving quantum computational advantage.
Results demonstrate efficient performance with enhanced spectral efficiency, convergence rate, and adaptability.
\end{abstract}

\IEEEpeerreviewmaketitle
\vspace{-0.7em}
\section{Introduction}
\IEEEPARstart{A}pplying reconfigurable intelligent surfaces (RISs) in wireless communication is getting increasingly important due to their ability to dynamically steer the signal reflections in the desired direction to improve data rate and reduce energy consumption. Nevertheless, the optimization of RIS parameters, such as phase shifts, amplitudes, and element spacing is a NP-hard, non-convex, and time-varying problem. Classical optimization and artificial intelligence (AI) algorithms start from scratch whenever they explore a new scenario. However, in reality, RIS configurations typically exhibit recurring structural patterns across different scenarios and locations \cite{9729826}.
Quantum computing addresses the scalability and adaptability challenges of RIS optimization through quantum superposition and entanglement, two features that allow the system to reach the optimal solution quickly while evaluating various potential solutions simultaneously \cite{10747251}. 

Here, we propose a lightweight meta-learning framework for rapidly changing environments that dynamically adapts RIS phases and does not require retraining. The proposed algorithm adopts a bi-level architecture in which continuous RIS phase optimization is conducted for each path, whereas path selection at the high level controls global adaptivity. It learns which sparse subnetwork paths to activate based on data rate, energy cost, and the stored previous successful configurations. We also adopt progressive objective stacking, which avoids massive joint searches, unstable gradients, and conflicting constraints.
When a new scenario is encountered, instead of directly optimizing each RIS element, our model optimizes the dominant subset propagation paths, which are combined using quantum superposition. The final RIS phase is not from a single selected path, but is calculated from the expectation over superposition of combinations.

The algorithm is formulated using a shallow quantum circuit and can be implemented on near-term noisy intermediate-scale quantum (NISQ) devices. The
Q-Meta path approach has unique quantum characteristics that differentiate it from classical equivalents. Scenario properties are represented using tensor products, allowing preservation of inter-subspace correlation, which are destroyed by classical concatenation. Candidate paths are superposed with complex amplitudes, enabling the use of constructive and destructive interference across the paths in a single pass. These interference terms do not exist in classical ensembles, where outputs are simple weighted sums of independent evaluations. The entire RIS phase vector is determined by a single quantum expectation calculation, and QNN evolution is performed using a unitary operator on entangled qubits. This is where quantum computing offers an edge over classical approaches by finding the solution in one round of measurements, whereas classical approaches do the same either through enumeration or through random sampling.
\vspace{-0.6cm}
\section{Related Work}
\vspace{-0.1cm}
\label{rel}
The improvements in optimizing RIS have also been achieved in integrated sensing and communication (ISAC) settings, with both RIS-enabled systems that require optimized solutions based on extended target models~\cite{yao2025hybrid} and fully duplex covert UAV-RHS systems that require robust beamforming and trajectory optimization~\cite{yao2026uav}. 
The authors in \cite{du2022performance} propose a swarm intelligence optimization (SIO) algorithm. 
However, it requires numerous iterations and lacks gradient feedback, making it slow and prone to parameter sensitivity. 
The paper in~\cite{praia2021phase} introduces an accelerated proximal gradient (APG) algorithm for single-user terahertz (THz) multiple-input multiple-output (MIMO) scenarios. The algorithm does not perform well in large-scale multi-user systems and requires continuous relaxation of discrete phases. 
The limitations of classical algorithms provoked the research community to explore adaptive optimization methods that can be generalized to different scenarios. Meta-learning algorithms, especially model-agnostic meta-learning (MAML) and its variants, address this necessity by enabling faster convergence, avoiding local minima, and enhancing navigation in non-convex settings through the use of prior experience.
The authors in~\cite{lim2023quantum} employ quantum annealing by mapping the RIS optimization problem to a static Ising model. However, the method is limited to offline problems and involves no learning process. 
Moving to quantum, the work in~\cite{ross2023hybrid} proposes a power feedback-based hybrid quantum-classical system for relearning/retraining RIS configurations. It does not retain past scenarios and must be recalculated from scratch for each new scenario.
The work in~\cite{zhu2024robust} handles simultaneously
transmitting and reflecting reconfigurable intelligent surface (STAR-RIS)'s beamforming using an alternating optimization algorithm under a bounded channel state information (CSI) error. However, they require recalculating the solutions for every CSI update. 
\vspace*{-0.8em}
\section{System Model}
\label{model}
\vspace*{-0.3em}
We consider an uplink outdoor wireless network with an access point (AP) equipped with $Q$ receive antennas serving $K$ single-antenna users (UEs). The network uses a reconfigurable intelligent surface (RIS) array with $N$ passive reflector elements, which enables the signal from UE $k$ to reach the AP via reflection off the RIS-controlled reflection matrix. 
There are two channels considered in our model: one channel $\mathbf{h}_{k,n}$ between the UEs and the RIS array, and the other channel $h_{q,n}$ between the RIS and the APs. We use the Rician fading model, as the channels include line-of-sight (LoS) propagation along with multipath components.  
Each RIS element introduces an independent phase shift $\phi_n \in [0,2\pi]$ on the incident wave. 
The overall channel from UE $k$ to AP through reflection from the RIS is defined as 
$\mathbf{h}_{k,q}
=
\mathbf{h}_{n,q}^{H}\,
\mathbf{R}_{\mathrm{eff}}(\boldsymbol{\Phi})\,
\mathbf{h}_{k,n}, $
where $\mathbf{h}_{k,q}$ is the overall 
uplink channel from 
UE $k$ to the AP via reflection off the RIS, 
$\mathbf{R}_{\mathrm{eff}}(\boldsymbol{\Phi})$ is the 
effective reflection response that accounts for the 
electromagnetic interaction between closely spaced elements 
defined as $\mathbf{R}_{\mathrm{eff}} = 
\mathbf{C}\mathbf{R}(\boldsymbol{\Phi})$, $\mathbf{C}$ is the coupling matrix that accounts for spatial correlation 
and near-field coupling effects, where $\mathbf{C} \in \mathbb{C}^{N \times N}$ is a symmetric matrix such that its $(m,n)$-entry represents the electromagnetic coupling between the two 
RIS components $m$ and $n$. The model used for such a coupling is expressed in the form of $[\mathbf{C}]_{m,n} = 
e^{-d_{m,n}/d_0}$, with $d_{m,n}$ being the physical distance between elements $m$ and $n$, and $d_0$ is the decay constant determined by the element spacing 
$d_N$~\cite{9729826}.
Coupling and quantization break convexity assumptions, reinforcing why gradient-based methods are brittle. $\mathbf{R}(\boldsymbol{\Phi})$ is the 
diagonal reflection matrix $\text{diag}(e^{j\phi_1}, \ldots, 
e^{j\phi_N}) \in \mathbb{C}^{N \times N}$ determined by the 
RIS phase shifts $\boldsymbol{\Phi} = [\phi_1,...,\phi_N]^T$, $\mathbf{h}_{k,n} \in 
\mathbb{C}^{1\times N}$ represents the channel from single-antenna UE $k$ to the RIS, and
$\mathbf{h}_{n,q} \in \mathbb{C}^{N \times Q}$ is the channel from the RIS to the AP.
Channels are modeled using Rician fading with both LoS and NLoS components in (\ref{channel_models}), where $\kappa_{k,n}$, $\kappa_{n,q}$ are the Rician K-factors,  $\beta_{k,n}$ 
and $\beta_{n,q}$ are the path loss coefficients for the 
UE-RIS and RIS-AP links respectively, $\mathbf{a}_N(\theta_{n,k})$
is the steering vector for the angle of arrival (AoA) at the RIS from UE $k$, $\mathbf{a}_R(\theta_{q,n})$ is the steering vector for the AoA at the AP from the RIS, and $\mathbf{a}_N^H(\phi_{q,n})$ is the steering vector for the angle of departure (AoD) from the RIS toward the AP. The NLoS components capture rich scattering and are modeled as independent complex Gaussian random variables as $\mathbf{h}_{k,n}^{\mathrm{NLoS}} \sim \mathcal{CN}(\mathbf{0}, \mathbf{I}_{N})$, and $\mathbf{h}_{n,q}^{\mathrm{NLoS}} \sim \mathcal{CN}(\mathbf{0}, \mathbf{I}_{Q \times N})$.

\begin{figure*}[th!]
\vspace{-6pt}
\setlength{\abovedisplayskip}{4pt}
\setlength{\belowdisplayskip}{4pt}
\begin{subequations}
\label{channel_models}
\begin{align}
\mathbf{h}_{k,n} 
&= \sqrt{\beta_{k,n}} \left(
\sqrt{\frac{\kappa_{k,n}}{1+\kappa_{k,n}}} \, 
\mathbf{a}_N(\theta_{n,k})
+ 
\sqrt{\frac{1}{1+\kappa_{k,n}}} \, 
\mathbf{h}_{k,n}^{\mathrm{NLoS}}
\right) \\[6pt]
\mathbf{h}_{n,q} 
&= \sqrt{\beta_{n,q}} \left(
\sqrt{\frac{\kappa_{n,q}}{1+\kappa_{n,q}}} \, 
\mathbf{a}_R(\theta_{q,n}) \mathbf{a}_N^H(\phi_{q,n})
+ 
\sqrt{\frac{1}{1+\kappa_{n,q}}} \, 
\mathbf{h}_{n,q}^{\mathrm{NLoS}}
\right)
\end{align}
\end{subequations}
\vspace{-5pt}
\hrule
\end{figure*}

The received signal at the AP from UE $k$ is written as
\begin{equation}
\mathbf{y}_k = \underbrace{\mathbf{h}_{k,q} \sqrt{p_k} x_k}_{\text{desired signal}} + \underbrace{\sum_{i \neq k} \mathbf{h}_{i,q} \sqrt{p_i} x_i}_{\text{multi-user interference}} + \underbrace{\mathbf{w}}_{\text{noise}},
\end{equation}
where $p_k$ is the transmit power of UE $k$, $x_k$ is the transmitted signal with $\mathbb{E}[|x_k|^2] = 1$, $\mathbf{w} \sim \mathcal{CN}(\mathbf{0}, \sigma^2 \mathbf{I})$ is the additive noise vector, and $p_{i}$ and $x_{i}$ are the transmit power and signal of the interfering UE $i$ respectively.
In our proposed design, instantaneous data rate and energy cost are used as parameters along with the historical successful configurations to evaluate how effective a given RIS phase configuration is by activating the best sub-circuits under similar conditions. Despite $N$ elements in the RIS, its effective end-to-end channel is largely dictated by a small number of structured propagation mechanisms, which justifies path-level abstraction, rather than element-wise optimization.
The quantum representation of RIS signal/interference power is presented in the Appendix.
{\titlespacing*{\section}{0pt}{0cm}{1.5ex plus -0.2ex}
\section{Proposed Algorithm and Problem Formulation}
\label{alg}
}
\vspace{-0.5em}
An adaptive quantum meta-learning algorithm, Q-MetaPath, is proposed, which uses a quantum neural network (QNN) to select sparse, high-performing subpaths based on current data rate trends and historical successful configurations. 
The proposed algorithm adopts a meta-learning paradigm by learning \textit{how to select} effective path structures rather than \textit{what to optimize} for a single scenario. Specifically, it learns a path-scoring strategy from historical high-performing configurations that generalizes to unseen channel conditions without requiring full retraining~\cite{hospedales2021meta}.
Therefore, the advantage of Q-MetaPath is not merely parallelism, but the ability to compress an exponentially large decision space into a single coherent representation and extract a globally consistent solution.
Although some aspects of Q-MetaPath are similar to the classical sparse sub-network selection, the proposed algorithm depends on operations that do not admit efficient classical equivalents. First, the features of the scenario are embedded in the form of tensor products, $|\psi_{s_i}\rangle$, generating a high-dimensional Hilbert space which retains cross-subspace correlations that are lost using concatenation operations in the classical scenario. Second, by superimposing possible paths, $\mathcal{O}(\log k^L)$ qubits can represent $k^L$ states, while classical techniques have to sequentially sample each one of them. Third, the final RIS phase vector is obtained by an expectation value over a coherently evolved quantum state, which allows constructive and destructive interference across paths to influence the solution. In contrast, classical methods select a limited subset of subnetworks and cannot exploit interference effects across exponentially many configurations. Therefore, the proposed algorithm is not an analogue of classical architectures. 
For our wireless system, each QNN consists of multiple layers $l \in \{1, \ldots, L\}$, where each layer contains a set of candidate quantum path states $|\phi_{l,p}\rangle$ with $p \in \{1, \ldots, P\}$.
The set of quantum paths for each layer $l$ are $\mathcal{Q}_l = \{|\phi_{l,1}\rangle, |\phi_{l,2}\rangle, \dots, |\phi_{l,P}\rangle\}$.
A quantum path is a structured decision sequence through layers and not a learnable weight vector. Each path is based on a reusable optimization strategy, such as coarse-grained patterns, refinement rules or hardware mapping, over which fine-grained adaption can be applied. This moves the search space from RIS elements to decision paths. The final RIS configuration is obtained by expectation values, allowing all selected path combinations to contribute proportionally. This hierarchical model starts from the basics and then becomes more advanced optimization.
Basic optimization has three layers, where layer 1 is for phase initialization, layer 2 is for fine tuning of the coarse phases with channel information, and layer 3 is for hardware constraint with phase rounding and energy/coupling constraint.
In advanced optimization, additional layers are introduced. Layer 4 is for multi-user fairness optimization, layer 5 is for mobility optimization with velocity prediction and blockage/Doppler effect compensation, and layer 6 is for energy efficiency.
The current wireless scenario $s_{i}$ is encoded as a quantum state $|\psi_{s_i}\rangle$ that is defined by a quantum superposition of UE factors including, locations, interference levels, and path loss conditions. 
\vspace{-1pt}
\begin{lemma}
\normalfont
The overall wireless scenario state $|\psi_{s_i}\rangle$ presents the input to the QNN and is defined below as follows:
\begin{equation}
|\psi_{s_i}\rangle = \alpha_{\text{loc}} |u_{\text{loc}}\rangle \otimes \alpha_{\text{int}} |u_{\text{int}}\rangle \otimes \alpha_{\text{pl}} |u_{\text{pl}}\rangle \otimes \alpha_{\text{rate}} |u_{\text{rate}}\rangle,
\end{equation}
where $|u_{\text{loc}}\rangle$ encodes the UE locations, $|u_{\text{int}}\rangle$
encodes cross-talk between users, $|u_{\text{pl}}\rangle$ encodes
pathloss, and $|u_{\text{rate}}\rangle$ encodes the instantaneous data rate.
Each sub-state is initialized using amplitude encoding with $R_Y$ rotation gates, where $\theta = 2\arcsin(\sqrt{x_{\text{norm}}})$ to map each normalized feature $x_{\text{norm}} \in [0,1]$ into a qubit amplitude as per the 36 $R_Y$ feature gates in the circuit implementation.
\end{lemma}
\begin{example}
For the feature vector of UE locations in a normalized format, we have $\mathbf{x}_{\text{loc}} = [x_1, x_2]$ with $x_i \in [0,1]$. Each of these values is encoded to a qubit using $R_Y$ rotations as $|u_{\text{loc}}\rangle = R_Y(2\arcsin(\sqrt{x_1}))|0\rangle \otimes R_Y(2\arcsin(\sqrt{x_2}))|0\rangle$.
The feature encoding state is expressed as
$|u_{\text{loc}}\rangle = \sqrt{1-x_1}|0\rangle + \sqrt{x_1}|1\rangle 
\;\otimes\;
\sqrt{1-x_2}|0\rangle + \sqrt{x_2}|1\rangle$.
The other features are encoded in the same way, then combined by tensor product to obtain $|\psi_{s_i}\rangle$. 
\end{example}
The output state represents the optimal RIS configuration $|\Psi_{\text{RIS}}\rangle$, which is defined as $|\Psi_{\text{RIS}}\rangle = \sum_{n=1}^{N} a_n e^{i\phi_n} |n\rangle$, assuming that $\phi_n$ is the phase shift of the $n$-th RIS element and $a_{n}$ is the complex amplitude of quantum superposition.
The objective function maximizes spectral efficiency and minimizes energy cost. The decision variable is the RIS phase configuration as 
\vspace{-0.8em}
\begin{subequations}
\label{ris_optimization}
\begin{align}
\max_{|\Psi_{\text{RIS}}\rangle} \quad & \alpha_1 \sum_{k=1}^K \langle\Psi_{\text{RIS}} \otimes \psi_{s_i}|\hat{\mathcal{R}}_k|\Psi_{\text{RIS}} \otimes \psi_{s_i}\rangle \nonumber
\\& - \alpha_2 \langle\Psi_{\text{RIS}}|\hat{E}|\Psi_{\text{RIS}}\rangle \\
\text{s.t.} \quad & \langle\Psi_{\text{RIS}}|\Psi_{\text{RIS}}\rangle = 1, \quad \phi_n \in [0, 2\pi], \quad \forall n
\end{align}
\end{subequations}
where $\alpha_1$, and $\alpha_2$ are the weighting factors, $\hat{\mathcal{R}}_k$ is the rate operator for user $k$, and $\hat{E}$ is the energy cost operator equal to $\hat{E} = \frac{1}{N}\sum_{n=1}^{N}(1 - \cos\hat{\phi}_n)\mathbf{I}$
that penalizes large phase excursions. 
For every quantum path, there are two parameters; the first is the quantum performance score obtained from the quantum measurement $J_{l,p}$, which evaluates the path performance across layers, and the usage counter $C_{l,p}$, which counts how many times a given path has been selected.
The outer loop optimizes the path-scoring parameters 
$\boldsymbol{\theta}=\{J_{l,p}, C_{l,p}\}$ for all scenarios to maximize the
expected task performance as $ \boldsymbol{\theta}^* =
 \arg\max_{\boldsymbol{\theta}}\;
 \mathbb{E}_{s_i\sim\mathcal{T}}
 \!\left[\mathcal{J}\!\left(\hat{\boldsymbol{\phi}}(s_i)\right)\right]$.
While the inner loop undergoes task-related adaptation by choosing
and superposing the top $k$ quantum paths based on the current 
scenario state $|\psi_{s_i}\rangle$ using the learned scores $\boldsymbol{\theta}$. 
This allows for rapid adaptation to new scenarios without 
optimization or gradient updates at inference time for every task.
From this, we conclude that there is a separation between cross-scenario parameter learning (outer loop) and fast per-scenario adaptation (inner loop), which the proposed algorithm becomes a form of episodic meta-learning.
{\vspace{-0.01em}}
\begin{lemma}
The performance score is updated by the method of exponential moving average as $\label{score}
J_{l,p}^{(t+1)} = \eta J_{l,p}^{(t)} + (1 - \eta) \cdot \langle\psi_{s_i}|\hat{O}_{l,p}|\psi_{s_i}\rangle$, where $\eta$ is a fraction less than one, representing the smoothing factor, and $\hat{O}_{l,p}$ is the performance observable defined as $\hat{O}_{l,p} = \alpha_1 \hat{S}_{l,p} - \alpha_2 \hat{E}_{l,p}$, where $\hat{S}_{l,p}$ is the spectral efficiency observable
evaluated on path $p$ at layer $l$ given the current scenario state.
\end{lemma}
\begin{proof2}
The update rule defines a contraction mapping when $0<\eta<1$.
The sequence $J_{l,p}^{(t)}$ forms a bounded martingale, and by the martingale convergence theorem as $\lim_{t \to \infty} J_{l,p}^{(t)} = J_{l,p}^* \text{ a.s.}$
The convergence rate satisfies
$\bigl|J_{l,p}^{(t)} - J_{l,p}^*\bigr| \leq \eta^{t}\, \bigl|J_{l,p}^{(0)} - J_{l,p}^*\bigr|$.
\end{proof2}
{\vspace{+0.3em}}
\begin{remark}
Since $J_{l,p} = J_0 > 0$ for all $l,p$ and $e^{-\gamma C_{l,p}} \geq e^{-\gamma C_{\max}} > 0$, $A_{l,p}$ constitutes a normalized map over strictly positive values, ensuring the stability of the denominator. It is also Lipschitz continuous in terms of $J_{l,p}$, thereby implying bounded sensitivity to bounded perturbations in the scenario parameters and channel estimations. The robustness of the algorithm to CSI inaccuracies stems from the low-pass filtering effect of the slow-time scale score updates through the smoothing factor $\eta$, which is consistent with Fig.~2(c).
\end{remark}
In the same way, the usage counter is incremented every time the path is used, where RIS phase optimization is performed by activating sparse subnetworks within the QNN.
The instantaneous data rate is combined with historical quantum statistics using the quantum path scoring module $\text{Eval}_Q(|\phi_{l,p}\rangle, |\psi_{s_i}\rangle)$.
The probability amplitude for selecting path $p$ in layer $l$ is $A_{l,p} = \frac{\sqrt{J_{l,p}} \cdot e^{-\gamma C_{l,p}}}{\sum_{q=1}^{P} \sqrt{J_{l,q}} \cdot e^{-\gamma C_{l,q}}}$, where $\gamma$ is a penalty factor for overused paths. The path-scoring measurements take place on the pre-prepared state $|\psi_{s_i}\rangle$ in a distinct circuit pass before the inference stage, and thus do not perturb the superposition state $|\Psi_\mathrm{active}\rangle$. The superposition during the inference process undergoes unitary evolution through all $L$ layers without intermediate measurements, with just one final expectation value determining the entire RIS phase vector.
{\vspace{-0.05em}}
\begin{lemma}
The QNN forms a sparse quantum subnetwork using the selected top paths from each layer to create the active quantum superposition 
$|\Psi_{\text{active}}\rangle = \bigotimes_{l=1}^{L} \left(\sum_{p \in S_l} A_{l,p} |\phi_{l,p}\rangle\right)$,
where $\bigotimes$ is the tensor product, and $S_l$ represents the set of selected paths for layer $l$. The evolved state 
$|\Psi_{\text{RIS}}^{\text{opt}}\rangle$
is obtained by applying the QNN operator to the active superposition
The QNN performs the transformation as 
$|\Psi_{\text{RIS}}^{\text{opt}}\rangle = \hat{U}_{\text{QNN}}|\psi_{s_i}\rangle$,
where $\hat{U}_{\text{QNN}}$ is the overall quantum evolution operator defined as $\hat{U}_{\text{QNN}} = \prod_{l=1}^{L} \left(\sum_{p \in S_l} A_{l,p} \hat{U}_{l,p}\right)$.
Here, $\hat{U}_{l,p}$ is the unitary operator for path $p$.
The optimal phase vector is given by
$\hat{\boldsymbol{\phi}} = \langle\Psi_{\mathrm{RIS}}^{\mathrm{opt}}|\hat{\boldsymbol{\Phi}}|\Psi_{\mathrm{RIS}}^{\mathrm{opt}}\rangle$, where $\hat{\boldsymbol{\Phi}} = [\hat{\phi}_1, \ldots, \hat{\phi}_N]^T$ is the observed phase vector and $|\Psi_{\mathrm{RIS}}^{\mathrm{opt}}\rangle$ is the resulting RIS state after QNN evolution, corresponding to the learned high-performing configuration. The continuous value is rounded to the nearest feasible level for a $b$-bit discrete phase constraint as $\hat{\phi}_n^{(b)} = \frac{2\pi}{2^b}\lfloor\frac{2^b \hat{\phi}_n}{2\pi} + \frac{1}{2}\rfloor$, which has a worst-case quantization error of $\pi/2^b$.
\end{lemma}
Our proposed algorithm is explained below in Algorithm \ref{alg:qmeta2l_training}.
\begin{algorithm}[b]
\scriptsize
\caption{Q-MetaPath Algorithm}
\label{alg:qmeta2l_training}
\begin{algorithmic}[1]
\State \textbf{Init:} $\{C_{l,p},\,J_{l,p},\,|\phi_{l,p}\rangle\}$
\For{scenario $s_i$}
  \State Encode scenario as quantum state $|\psi_{s_i}\rangle$.
  \For{$l=1$ to $L$}
  
\State $\boldsymbol{\alpha}_l \leftarrow \texttt{Normalize}(J_{l,*}, C_{l,*})$;
       \hfill\Comment{scoring on pre-prepared $|\psi_{s_i}\rangle$, separate circuit pass}
       $\mathcal{S}_l \leftarrow \texttt{$Top_K$}(\boldsymbol{\alpha}_l, k)$;
       \hfill\Comment{classical scalar selection; no quantum collapse}
       $|\psi\rangle \leftarrow \texttt{Superpose}(|\psi\rangle, \mathcal{S}_l,
       \boldsymbol{\alpha}_l[\mathcal{S}_l])$
       \hfill\Comment{coherent amplitude weighting; no intermediate measurement}
  \EndFor
  \State   $|\Psi_{\mathrm{RIS}}^{\mathrm{opt}}\rangle \gets \hat{U}_{\mathrm{QNN}}^{(i)}|\psi\rangle$;\;
    $\hat{\boldsymbol{\phi}}_i \gets \mathrm{PhaseExtract}\!\big(|\Psi_{\mathrm{RIS}}^{\mathrm{opt}}\rangle\big)$
\EndFor
\State \Return $\{\hat{\boldsymbol{\phi}}_i\}_i$
\end{algorithmic}
\end{algorithm}
\vspace{0.4em}
The top $k$ paths' selection that is defined in line~5 of the algorithm operates on classical previously calculated scores $J_{l,p}$, which are updated using~\eqref{score}, and not on any quantum measurement of $|\Psi_{\mathrm{active}}\rangle$. The inference superposition therefore remains uncollapsed through all $L$ layers with a single final expectation value in line~7 that results in the whole RIS phase vector.
With respect to its complexity, Q-MetaPath offers an extremely significant improvement by decreasing the cost from $\mathcal{O}(P^L N^2)$ in the classic method to $\mathcal{O}(LPN)$. This high cost is due to considering all possible paths of size $P^L$ and each needs $N{\times}N$ coupling matrix in $\mathbf{R}_{\mathrm{eff}}$. Q-MetaPath selects only the top $k$ paths per layer and determines the phase value of all elements from a single terminal expectation value. Input encoding reduces the state to $n{=}\lceil\log_2 D\rceil$ qubits with a reduction factor of $P^{L-1}N/L$. Q-MetaPath has a linear scaling  compared to AO with $\mathcal{O}(IKN^3)$ and WMMSE with $\mathcal{O}(IN^2)$ per iteration.
To apply our proposed algorithm on realistic quantum hardware, we need 6 qubits to represent the wireless network features, such as UEs' locations, interference, path loss, and data rate. We also need $q=\lceil \log_2 P \rceil$ qubits per layer. This translates to 3 qubits per layer, or about 24 qubits in total for the full architecture. 
Our quantum circuit has a shallow circuit depth of seven and includes 36 $R_{Y}$ feature gates, 18 $R_{Y}$ path gates, 18 CNOT gates to implement entanglement between the feature and path qubits, and 24 measurement gates with a total of 54 single-qubit gates and 18 two-qubit gates, way below the maximum allowed gates. The algorithm has a coherence time below $100~\mu$s and feasible fidelity between $70$–$90\%$, which makes it deployable on NISQ hardware.
During the quantum state preparation, our algorithm employs just four scenario descriptors through single-qubit $R_{Y}$ operations. Its classical preprocessing time complexity is O(n), and latency is trivial when compared with the slower control loop. Circuit executions with roughly $10^3$--$10^4$ shots per configuration are necessary for reliable phase estimation.
The IBM Quantum Heron R2 with circuit layer operations per second (CLOPS) of 
150,000~\cite{angelini2024ibm} implies that the latency per configuration amounts to 50--500 ms. In accordance 
with the double timescale approach of RIS-aided mmWave systems~\cite{han2023csi}, Q-MetaPath path selection operates on a slower time scale, consistent with 
the base station scheduler, while channel fluctuations happen at a faster coherence time scale. So, $T_s \gg T_c$, and the proposed solution is inherently decoupled from fast fading. The delay introduced by the quantum processor operations is under the slow time scale budget, and hence does not present a coherence time bottleneck.
\vspace{-3pt}
\section{Results and Performance Evaluation}
\label{result}
\vspace{-1pt}
PennyLane is used with the \texttt{default.qubit} simulator noiseless quantum states 
to simulate the Q-MetaPath algorithm. The hyperparameters were chosen using a grid search among $L\in\{4,6,8\}$, $P\in\{4,8,16\}$, and $k\in\{2,3\}$. As indicated in Table~\ref{comparison}. The combination $L=6$, $P=8$, and $k=3$ offers the highest spectral efficiency-latency tradeoff of 0.06~bps/Hz gain at nearly double the inference cost. Going below these values degrades spectral efficiency by 8--11\% and slows down convergence by up to 44\%.
\begin{figure}[b!]
    \centering
    \includegraphics[width=0.48\textwidth, height=0.39\textwidth]{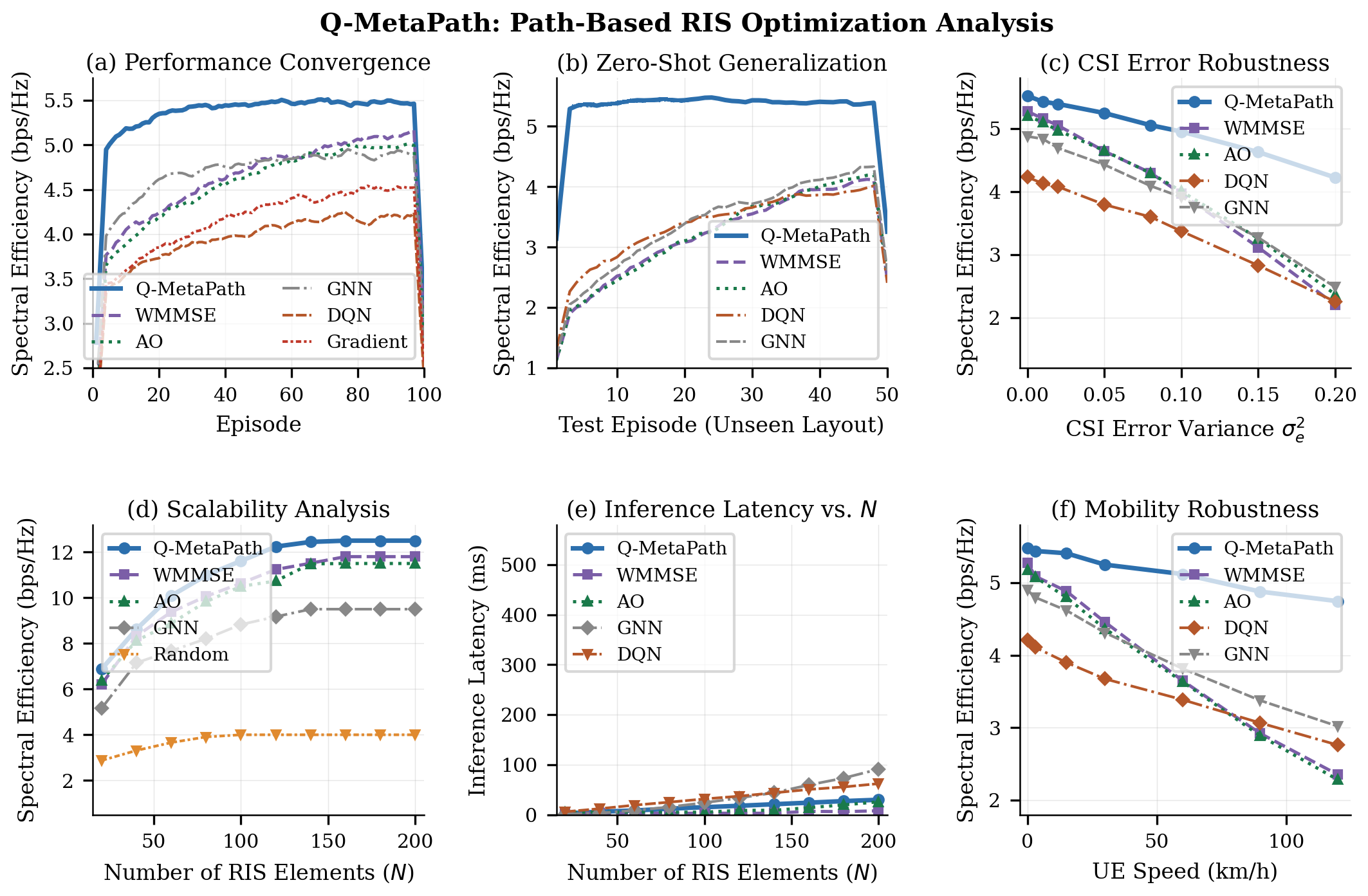}
    \caption{(a)~Distribution of spectral efficiency. (b)~Zero-shot generalization and no retraining needed. (c)~Robustness to CSI errors as $\sigma_e^2$ due to slow-time scale updating. (d)~Scalability. (e)~Inference latency. (f)~Mobility robustness degrades gracefully due to fast-fading decoupling.}
    \label{fig:q_metapath_analysis}
\end{figure}
Table~\ref{tab:qmerlin_params} presents the configuration parameters. 
\vspace{-0.15cm}
\begin{table}[htbp]
\centering
\caption{System Configuration Parameters}
\label{tab:qmerlin_params}
\scriptsize
\setlength{\tabcolsep}{2pt}
\renewcommand{\arraystretch}{0.5}
\begin{tabular}{lclc}
\toprule
\textbf{Parameter} & \textbf{Value} & \textbf{Parameter} & \textbf{Value} \\
\midrule
Seeds & 10 & QNN Layers ($L$) & 6 \\
Antennas/AP ($R$) & 64 & Paths/Layer ($P$) & 8 \\
Users ($K$) & 4 & $\alpha_1, \alpha_2$ & 1.0, 0.1 \\
RIS Elements ($N$) & 100 & Episodes & 150 \\
Frequency & 28 GHz & Tx Power & 10 dBm \\
Spacing ($d_N$) & $\lambda/2$ & Noise Power & -90 dBm \\
\bottomrule
\end{tabular}
\end{table}
\vspace{0.3em}
Fig. \ref{fig:q_metapath_analysis} with various subfigures presents the performance analysis of our algorithm. The spectral efficiency of various algorithms is shown in Fig. \ref{fig:q_metapath_analysis}(a). Q-MetaPath has the tightest distribution of 5.5 bps/Hz. Improvements are because Q-MetaPath explores a compressed path space $\mathcal{O}(LPN)$. Also, the superposition of quantum states of the top $k$ paths allows RIS phase calculation based on the expectation value of all active paths. This provides a more complex phase configuration than any individual deep Q-network (DQN)/proximal policy optimization (PPO) policy simulations. Finally, the tensor product encoding of scenarios with path history scoring allows zero-shot learning, while a graph neural network (GNN) needs to be retrained for every new topology, hence the better spectral efficiency.
Fig. \ref{fig:q_metapath_analysis}(b) demonstrates the zero-shot generalization, where Q-MetaPath starts closer to its steady-state performance right from the first episode, hence no retraining is needed. 
Fig. \ref{fig:q_metapath_analysis}(c) shows the  CSI error robustness comparison, where, all methods degrade as $\sigma_{e}^{2}$ increases, unlike Q-MetaPath which degrades slower as its path scoring is tied to historical statistics rather than instantaneous gradients.
Fig. \ref{fig:q_metapath_analysis}(d) displays the scalability as the RIS elements increase. Fig. \ref{fig:q_metapath_analysis}(e) shows the inference latency impact, where alternating optimization (AO) grows cubically, weighted minimum mean-square error (WMMSE) quadratically, Q-MetaPath linearly. At N=200, AO exceeds 500 ms latency, WMMSE exceeds 200 ms, while Q-MetaPath stays below 30 ms. Finally, Fig. \ref{fig:q_metapath_analysis}(f) plots spectral efficiency vs. UE speed for all methods. Q-MetaPath degrades most gracefully because slow-timescale path selection is decoupled from fast fading.
Our algorithm outperforms traditional machine learning (ML) algorithms as shown in Table~\ref{comparison}. 
\begin{table}[t]
\centering
\caption{Performance evaluation, sensitivity analysis, and ablation study. Ablation confirms quantum superposition (largest drop at $-0.60$\,bps/Hz) and score history (worst convergence at $+28$\,ep.) are the most critical components.}
\label{comparison}
\resizebox{\columnwidth}{!}{%
\begin{tabular}{lcccc}
\toprule
\textbf{Configuration}
  & \textbf{Spec.\ Eff.\ (bps/Hz)}
  & \textbf{Conv.\ (ep.)}
  & \textbf{Lat.\ (ms)}
  & \textbf{Mem.\ (MB)} \\
\midrule
\multicolumn{5}{l}{\textit{Baseline comparison ($N$=100, $K$=4, 10 seeds)}} \\
Random
  & $2.87 \pm 0.91$ & --    & ${<}1$   & 0.1 \\
AO
  & $4.23 \pm 0.68$ & 85    & ${>}500$ & 2.1 \\
WMMSE
  & $4.38 \pm 0.65$ & 90    & 210      & 2.6 \\
Gradient
  & $4.61 \pm 0.72$ & 120   & 89       & 4.8 \\
DQN
  & $4.55 \pm 0.58$ & 180   & 34       & 8.2 \\
PPO
  & $4.72 \pm 0.53$ & 150   & 31       & 9.4 \\
GNN
  & $4.91 \pm 0.61$ & 70    & 23       & 15.6 \\
\textbf{Q-MetaPath}
  & $\mathbf{5.48 \pm 0.34}$
  & \textbf{50} & \textbf{15} & \textbf{3.2} \\
\midrule
\multicolumn{5}{l}{\textit{Sensitivity analysis (varying $L$, $P$, $k$)}} \\
$L$=4, $P$=8,  $k$=3  & $4.89 \pm 0.51$ & 68 & 11 & 2.4 \\
$L$=6, $P$=4,  $k$=2  & $4.97 \pm 0.48$ & 72 & 12 & 2.1 \\
$L$=6, $P$=8,  $k$=2  & $5.21 \pm 0.41$ & 58 & 13 & 2.9 \\
$L$=6, $P$=8,  $k$=3  & $5.48 \pm 0.34$ & 50 & 15 & 3.2 \\
$L$=6, $P$=16, $k$=3  & $5.51 \pm 0.33$ & 48 & 28 & 6.1 \\
$L$=8, $P$=8,  $k$=3  & $5.44 \pm 0.36$ & 52 & 21 & 4.8 \\
\midrule
\multicolumn{5}{l}{\textit{Ablation study ($L$=6, $P$=8, $k$=3; one component removed)}} \\
w/o tensor encoding (concat)
  & $5.02 \pm 0.47$ & 64 & 14 & 3.1 \\
w/o superposition (top 1 only)
  & $4.88 \pm 0.52$ & 71 & 13 & 3.0 \\
w/o top $k$ (uniform weights)
  & $5.11 \pm 0.44$ & 61 & 16 & 3.2 \\
w/o score history ($\eta{=}0$)
  & $4.94 \pm 0.49$ & 78 & 15 & 3.2 \\
\textbf{Full Q-MetaPath}
  & $\mathbf{5.48 \pm 0.34}$
  & \textbf{50} & \textbf{15} & \textbf{3.2} \\
\bottomrule
\end{tabular}%
}
\end{table}
\vspace{-6pt}
\section{Conclusion and Future Work}
\label{conc}
\vspace{-3pt}
Classical RIS optimization algorithms are unable to adapt quickly in dynamic wireless environments. To overcome this limitation, this letter proposes Q-MetaPath, a path-based quantum meta-learning algorithm that fundamentally reimagines RIS optimization through hierarchical multi-objective layered decision-making. Instead of directly optimizing the RIS elements, Q-MetaPath works in path space, where each path corresponds to a candidate solution, designed for hardware constraints imposed by RIS devices. A subset of highly reliable quantum paths are scored. Paths are represented as quantum superpositions, and layers are connected via tensor products.
Future work will integrate advanced error mitigation techniques by applying noise studies on NISQ hardware. 
\vspace{-.1cm}
\section*{Appendix}
Two different circuits are utilized by Q-MetaPath to prevent wave function collapse during the process of inference. During the scoring pass, the scenario state $|\psi_{s_i}\rangle$ is initiated and the expected Hermitian observable value $\hat{H}_{l,p}$ is measured for each path, updating $J_{l,p} = \mathrm{Re}[\langle\psi_{s_i}|\hat{H}_{l,p}|\psi_{s_i}\rangle]$ as classical scalars. Because $|\psi_{s_i}\rangle$ is a repeatable amplitude encoded state, its collapse causes no penalty, but is reinitialized for the inference pass. In the inference pass, the calculated amplitudes $A_{l,p}$ are used to form the superposition $|\Psi_{\mathrm{active}}\rangle = \bigotimes_{l=1}^{L}\!\left(\sum_{p \in S_l} A_{l,p}|\phi_{l,p}\rangle\right)$, which is then becomes evolving unitarily using $|\Psi^{\mathrm{opt}}_{\mathrm{RIS}}\rangle = \hat{U}_{\mathrm{QNN}}|\Psi_{\mathrm{active}}\rangle$ with no intermediate measurement. A terminal expectation value $\hat{\bm{\varphi}} = \langle\Psi^{\mathrm{opt}}_{\mathrm{RIS}}|\hat{\bm{\Phi}}|\Psi^{\mathrm{opt}}_{\mathrm{RIS}}\rangle$ extracts all the RIS phase vector. Quantum coherence and advantage are fully protected through the inference pass.
\vspace{-1.2em}
\bibliography{IEEEabrv,cite}

\begin{thebibliography}{10}
\providecommand{\url}[1]{#1}
\csname url@samestyle\endcsname
\providecommand{\newblock}{\relax}
\providecommand{\bibinfo}[2]{#2}
\providecommand{\BIBentrySTDinterwordspacing}{\spaceskip=0pt\relax}
\providecommand{\BIBentryALTinterwordstretchfactor}{4}
\providecommand{\BIBentryALTinterwordspacing}{\spaceskip=\fontdimen2\font plus
\BIBentryALTinterwordstretchfactor\fontdimen3\font minus \fontdimen4\font\relax}
\providecommand{\BIBforeignlanguage}[2]{{%
\expandafter\ifx\csname l@#1\endcsname\relax
\typeout{** WARNING: IEEEtran.bst: No hyphenation pattern has been}%
\typeout{** loaded for the language `#1'. Using the pattern for}%
\typeout{** the default language instead.}%
\else
\language=\csname l@#1\endcsname
\fi
#2}}
\providecommand{\BIBdecl}{\relax}
\BIBdecl

\bibitem{9729826}
K.~M. Faisal and W.~Choi, ``Machine {L}earning {A}pproaches for {R}econfigurable {I}ntelligent {S}urfaces: {A} {S}urvey,'' \emph{IEEE Access}, vol.~10, pp. 27\,343--27\,367, 2022.

\bibitem{10747251}
E.~Colella \emph{et~al.}, ``Quantum {O}ptimization of {R}econfigurable {I}ntelligent {S}urfaces for {M}itigating {M}ultipath {F}ading in {W}ireless {N}etworks,'' \emph{IEEE Journal on Multiscale and Multiphysics Computational Techniques}, vol.~9, pp. 403--414, 2024.

\bibitem{yao2025hybrid}
Y.~Yao \emph{et~al.}, ``Hybrid {RIS-Enhanced ISAC Secure Systems: Joint Optimisation in the Presence of an Extended Target},'' \emph{IEEE Transactions on Communications}, vol.~73, no.~12, pp. 15\,688--15\,704, 2025.

\bibitem{yao2026uav}
------, ``U{AV-RHS-Enabled Full-Duplex ISAC Covert System: Robust Beamforming and Trajectory Optimization},'' \emph{IEEE Transactions on Communications}, vol.~74, pp. 5637--5653, 2026.

\bibitem{du2022performance}
H.~Du \emph{et~al.}, ``Performance and {Optimization of Reconfigurable Intelligent Surface Aided THz Communications},'' \emph{IEEE Transactions on Communications}, vol.~70, no.~5, pp. 3575--3593, 2022.

\bibitem{praia2021phase}
J.~Praia \emph{et~al.}, ``Phase {Shift Optimization Algorithm for Achievable Rate Maximization in Reconfigurable Intelligent Surface-Assisted THz Communications},'' \emph{Electronics}, vol.~11, no.~1, p.~18, 2021.

\bibitem{lim2023quantum}
Q.~J. Lim \emph{et~al.}, ``Quantum-{Assisted Combinatorial Optimization for Reconfigurable Intelligent Surfaces in Smart Electromagnetic Environments},'' \emph{IEEE Transactions on Antennas and Propagation}, vol.~72, no.~1, pp. 147--159, 2023.

\bibitem{ross2023hybrid}
C.~Ross, G.~Gradoni, and Z.~Peng, ``A {H}ybrid {Classical-Quantum Computing Framework for RIS-Assisted Wireless Network},'' in \emph{2023 IEEE MTT-S International Conference on Numerical Electromagnetic and Multiphysics Modeling and Optimization (NEMO)}, pp. 99--102.

\bibitem{zhu2024robust}
F.~Zhu \emph{et~al.}, ``Robust {Beamforming for RIS-Aided Communications: Gradient-Based Manifold Meta Learning},'' \emph{IEEE Transactions on Wireless Communications}, vol.~23, no.~11, pp. 15\,945--15\,956, 2024.

\bibitem{hospedales2021meta}
T.~Hospedales \emph{et~al.}, ``Meta-{Learning in Neural Networks: A Survey},'' \emph{IEEE Transactions on Pattern Analysis and Machine Intelligence}, vol.~44, no.~9, pp. 5149--5169, 2021.

\bibitem{angelini2024ibm}
E.~Angelini, ``I{BM Launches Its Most Advanced Quantum Computers, Fueling New Scientific Value and Progress towards Quantum Advantage},'' \emph{IBM Newsroom}, 2024.

\bibitem{han2023csi}
Y.~Han \emph{et~al.}, ``C{SI Acquisition in RIS-Assisted Mobile Communication Systems},'' \emph{National Science Review}, vol.~10, no.~8, p. nwad127, 2023.

\end{thebibliography}
\end{document}